# Influence of Magnetic Anisotropy on Laser-induced Precession of Magnetization in Ferromagnetic Semiconductor (Ga,Mn)As

Tesařová Naďa[1], Rozkotová Eva[1], Reichlová Helena[1], Malý Petr[1], Novák Vít[2], Cukr Miroslav[2], Jungwirth Tomáš[2], and Němec Petr*[1]

[1] *Faculty of Mathematics and Physics, Charles University in Prague, Ke Karlovu 3, 121 06, Prague, Czech Republic*
[2] *Institute of Physics ASCR v.v.i., Cukrovarnická 10, 162 53 Prague, Czech Republic*

The laser-induced precession of magnetization in (Ga,Mn)As samples with different magnetic anisotropy was studied by the time-resolved magneto-optical method. We observed that the dependence of the precession amplitude on the external magnetic field depends strongly on the magnetic anisotropy of (Ga,Mn)As and we explain this phenomenon in terms of competing cubic and uniaxial anisotropies. We also show that the corresponding anisotropy fields can be deduced from the magnetic field dependence of the precession frequency.

In the last decade, the diluted magnetic semiconductor (Ga,Mn)As attracted significant attention due to the fact that in this material the ferromagnetic order originates from the interaction between itinerant holes and localized Mn moments.[1, 2] The magnetic properties are, therefore, sensitive to the carrier concentration that make this material a rather interesting model system for the basic research.[3] In particular, the magnitude and/or the orientation of magnetization can be modified on the picosecond time scale.[4, 5]

In 2005, it was revealed that the optical excitation of (Ga,Mn)As by femtosecond laser pulses triggers a precession of ferromagnetically coupled Mn spins.[5] Since then, the photo-induced precession of magnetization has been investigated by several groups.[6-10] It was concluded that the precession of magnetization is a consequence of the laser pulse-induced change of the magnetic anisotropy. But the exact mechanism is still under debate.[6-12] Recently, we have shown that the magnetic anisotropy can be modified by laser pulses not only due to the temperature increase but also due to the increase of holes concentration.[11] The precession of magnetization was studied in various (Ga,Mn)As samples and the influence of the Mn doping[9, 10] and of the thermal annealing[8, 9] was investigated. Nevertheless, the role of a magnetic anisotropy on the precession amplitude was not addressed in detail up to now.

The interpretation of the experimental results reported by different groups is significantly complicated by the fact that (Ga,Mn)As is, in principle, a disordered material. Therefore, a special care has to be taken when generalizing any phenomenon obtained in one particular sample to the universal behavior of this material system. Recently, we have reported on a systematic study of optical and magneto-optical properties of optimized set of (Ga,Mn)As epilayers spanning the wide range of accessible substitutional $Mn_{Ga}$ dopings.[13] The optimization of the materials in the series, which is performed individually for each nominal doping, minimizes the uncertainties in the sample parameters and produces high quality epilayers which are as close as possible to uniform uncompensated (Ga,Mn)As mixed crystals. For each nominal Mn doping $x$, the growth and post-growth annealing conditions were separately optimized in order to achieve the highest Curie temperature $T_c$ attainable at the particular $x$.[13] In this paper we report on a detailed study of the laser induced-precession of magnetization in two optimized (Ga,Mn)As samples that have a markedly different magnetic anisotropy. In particular, we show that the dependence of the precession amplitude on the external magnetic field depends strongly on the magnetic anisotropy and we explain this phenomenon in terms of competing cubic and uniaxial anisotropies in the samples.

---

*Corresponding author: nemec@karlov.mff.cuni.cz

The experiments were performed using two optimized 20 nm thick $Ga_{1-x}Mn_xAs$ epilayers with a distinct Mn content. The samples were grown on a GaAs(001) substrate by low-temperature molecular beam epitaxy and annealed in air. The nominal Mn doping ($x$), Curie temperature ($T_C$) and equilibrium hole concentration ($p_0$) are 3%, 77 K and $6.6 \times 10^{20}$ cm$^{-3}$ for sample $A$, and 7%, 159 K and $1.4 \times 10^{21}$ cm$^{-3}$ for sample $B$, respectively.[13] The magnetic anisotropy of the samples was studied by the superconducting quantum interference device (SQUID). The laser pulse-induced dynamics of magnetization was studied by the standard time-resolved magneto-optical (MO) technique.[8] As a light source, we used the Ti:sapphire laser that was tuned above the material bandgap ($h\nu = 1.64$ eV) and that produced laser pulses with a repetition rate of 82 MHz and a pulse width $\approx 200$ fs. The polarization of the pump pulses was circular (with a helicity controlled by the wave plate) and the probe pulses were linearly polarized along the [100] crystallographic direction. The energy fluence of the pump pulse was $\sim 28$ $\mu$Jcm$^{-2}$, with the pump to probe intensity ratio 20:1. The time-resolved MO data reported here correspond to the polarization-independent part of the pump-induced rotation of probe polarization, which was computed from the as-measured data by averaging the signals obtained for the opposite helicities of circularly polarized pump pulses.[11] The measured magneto-optical signals are due to the polar Kerr effect (PKE), which is sensitive to the out-of-plane component of magnetization, and magnetic linear dichroism (MLD), which is sensitive to the in-plane component of magnetization.[11] The samples were placed in a cryostat at the temperature of about 15 K. The external magnetic field $H_{ext}$ ranging from $\approx 0$ mT (i.e., smaller than 50 $\mu$T) to 550 mT was applied in the [010] crystallographic direction. In all time-resolved experiments we first applied $H_{ext} = 550$ mT and than we reduced the field to the required value. This initialization procedure was used to set the magnetization in the easy axis position that is the closest to the [010] crystallographic direction.

In order to investigate the influence of the magnetic anisotropy on the magnetization precession in $Ga_{1-x}Mn_xAs$ we focused on two samples with a different Mn content because the concentration of Mn atoms is one of the crucial parameters that determine the magnetic anisotropy in this material. Under equilibrium conditions, the magnetization is oriented in the easy axis direction, i.e. in the crystallographic direction that corresponds to the minimal energy of the system if no external magnetic field is applied. The position of the easy axis is given by the overall magnetic anisotropy that is quite complex in (Ga,Mn)As. It consists of two competing contributions. The first one is the biaxial [100] / [010] anisotropy which originates from the cubic symmetry of the GaAs host lattice. The second one is the growth-induced uniaxial anisotropy.[14] These anisotropies are strongly dependent on the carrier concentration and on the lattice temperature.[15] The investigated samples were grown on GaAs substrate and, consequently, they are compressively strained that resulted in the in-plane position of the easy axis.[2, 15]

In Fig. 1(a) we show the measured temperature dependent magnetization projections to [100], [-110] and [110] in-plane crystallographic directions in sample $A$. From the data it is clearly apparent that in the low temperature region (from 5 to $\approx 20$ K) the cubic anisotropy dominates, because the magnetization exhibits a maximal projection along the [100] direction. However, the contribution of the uniaxial anisotropy is not negligible as can be seen from the non-equal projections along the directions [-110] and [110]. This shows that at low temperatures the easy axis position is slightly tilted from the [100] direction towards the [-110] direction. With the increasing temperature, the cubic anisotropy is quenched faster than the uniaxial one[15] and, consequently, above ≈ 30 K the uniaxial anisotropy prevails. On the other hand, in sample *B* the uniaxial anisotropy dominates even at low temperatures that is clearly apparent from the small magnetization projection along the [110] direction. (We note that in sample *B* the temperature-induced quenching of magnetic anisotropies was so strong that above ≈ 100 K the measured data were strongly influenced by the magnetic field of 2 mT that was present in the SQUID during the measurement. Consequently, above ≈ 100 K the data could not be used for a reliable determination of the magnetic anisotropy in this sample.) In Fig. 1(c) and 1(d) we show the schematic illustrations of the four equivalent easy axis positions with respect to the main crystallographic directions in the investigated (Ga,Mn)As samples at 15 K. The observed magnetic anisotropies are fully consistent with the reported experimental[16] and theoretical[15] results and confirm the expected relative enhancement of the uniaxial anisotropy with the increasing Mn doping.

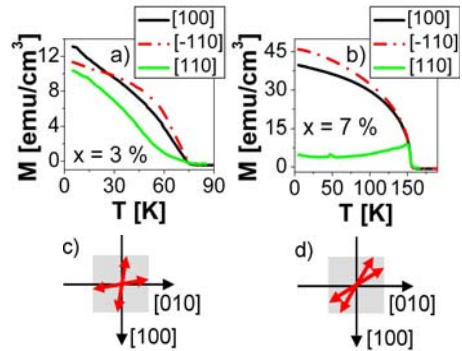

Fig. 1: Temperature dependence of magnetization projections to different crystallographic directions measured by SQUID in $Ga_{1-x}Mn_xAs$ samples with (a) x = 0.03 and (b) x = 0.07. (c) and (d) Schematic illustrations of four equivalent magnetization easy axis positions (red arrows) with respect to the main crystallographic directions [100] and [010] at 15 K for samples with x = 0.03 and x = 0.07, respectively.

In Fig. 2 we show the measured MO signal that is induced by the impact of pump pulses in sample *A* – the oscillatory signal is a signature of the laser-induced precession of magnetization.[6-10] The data can be fitted well by the exponentially damped harmonic function (see Fig. 2b) that is superimposed on a pulse-like background (see Fig. 2c)[8, 12]:

$$MO(t) = A\cos(\omega t + \Delta)e^{-t/\tau_G} + Ce^{-t/\tau_p} \quad (1)$$

where *A* and *C* are the amplitudes of the oscillatory and pulse-like function, respectively, *ω* is the angular frequency of precessing Mn spins, *Δ* is a phase factor, $\tau_G$ is the Gilbert damping time, and $\tau_p$ is the damping time of the pulse-like background. This analytical formula describes the phenomenological model,[11] where the signal is decomposed into a precession of magnetization around the quasi-equilibrium easy axis and into the laser-induced tilt of the easy axis. The oscillatory signal is strongly dependent on the external magnetic field $H_{ext}$. In Fig. 2 we show the data measured in sample *A* for two different magnetic fields applied along the [010] direction. In this sample, the external magnetic field suppresses the precession amplitude – see Fig. 2(b). On the contrary, as illustrated in Fig. 3, the precession amplitude can be strongly enhanced by $H_{ext}$ in sample *B*. The markedly different response of the precession amplitude to $H_{ext}$ in samples *A* and *B* is clearly apparent also from Fig. 4(a). We should note here, that the application of $H_{ext}$ influences also the equilibrium position of magnetization in the sample plane. Consequently, the magneto-optical response of the material can be changed due to the modification of the MLD magneto-optical coefficient, which depends on the angle between the magnetization and the linear polarization of the probe pulses.[11] This effect is negligible in sample A because $H_{ext}$ is applied rather close to the easy axis and thus the equilibrium position of magnetization does not change much with $H_{ext}$.

In sample *B*, the equilibrium position of magnetization is shifted monotonously from the easy axis towards the [010] direction [see Fig. 1(d)] with increasing $H_{ext}$. And even though this effect alters the MO response of the sample *B* it could not be the origin of the observed non-monotonous dependence of the precession amplitude on $H_{ext}$.

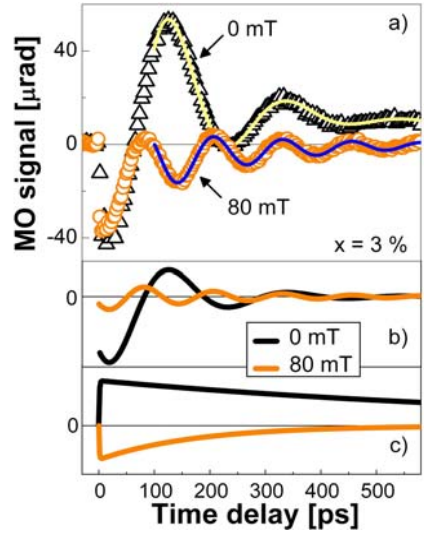

Fig. 2: (a) Time-resolved magneto-optical (MO) signal measured at 15 K in $Ga_{1-x}Mn_xAs$ sample with x = 0.03 for 0 mT and 80 mT external magnetic field applied along the [010] crystallographic direction (points). Solid lines are the fits by Eq. (1) – i.e., by a sum of the exponentially damped harmonic function and the pulse-like function, which are plotted separately in (b) and (c), respectively. The parameters of the fits are: $A$ = 114 μrad, $τ_G$ = 117 ps, $ω$ = 29.4 GHz, $Δ$ = 222 °, $C$ = 18.5 μrad, $τ_p$ = 880 ps and $A$ = 20 μrad, $τ_G$ = 213 ps, $ω$ = 50.3 GHz, $Δ$ = 211 °, $C$ = -14 μrad, $τ_p$ = 175 ps for 0 mT and 80 mT, respectively.

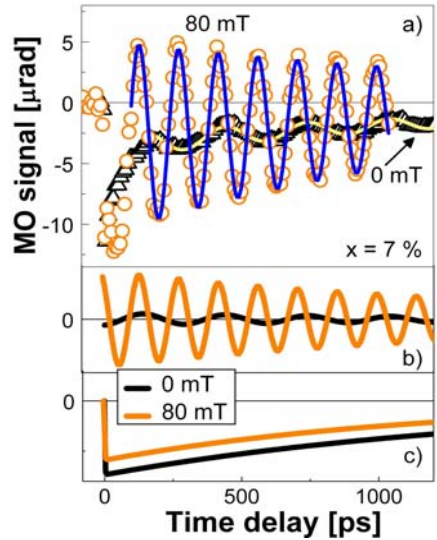

Fig. 3: Same as Fig. 2 for $Ga_{1-x}Mn_xAs$ sample with x = 0.07. The parameters of the fit are: $A$ = 1.0 μrad, $τ_G$ = 1560 ps, $ω$ = 21.0 GHz, $Δ$ = 280 °, $C$ = -3.6 μrad, $τ_p$ = 1500 ps and $A$ = 8.0 μrad, $τ_G$ = 1580 ps, $ω$ = 43.6 GHz, $Δ$ = 130 °, $C$ = -2.9 μrad, $τ_p$ = 1180 ps for 0 mT and 80 mT, respectively.

For the interpretation of the measured phenomena it is necessary to analyze the origin of the laser-induced precession of magnetization in detail. Absorption of a strong laser pulse in (Ga,Mn)As leads to a photogeneration of electron-hole pairs and to an increase of the sample temperature.[7, 9, 11] As already described above, both the cubic and the uniaxial anisotropies depend strongly on the concentration of holes and on the sample temperature but the dependences are rather different for them.[15] The impact of the pump pulse can modify the relative strength of these anisotropies and, therefore, it can change the position of the easy axis in the sample. This, in turn, triggers the precessional motion of magnetization which aims at pointing to this new quasi-equilibrium position where the energy of the system is minimal.[11] Nevertheless, the increase of the holes concentration and/or of the sample temperature does not always lead to the reorientation of the easy axis. For example, if the uniaxial anisotropy were far much stronger than the cubic anisotropy, the easy axis would be always located at [-110] crystallographic orientation. This shows that the laser-induced change of the easy axis position, which is triggering the precession of magnetization, can be observed only if the contribution of the cubic and the uniaxial anisotropies to the overall magnetic anisotropy in the sample are "comparable" (i.e., when the corresponding anisotropy fields contribute "similarly" to the free energy density $F$ in the sample[14]). The external magnetic field $H_{ext}$ appears as another uniaxial anisotropy in the sample (Zeeman term in $F$).[14] In our particular case, $H_{ext}$ is applied along the [010] crystallographic direction. Therefore, an increase of $H_{ext}$ modifies $F$ in such a way that it deepens the energy minimum connected with the cubic anisotropy which is located along the [010] direction. We also recall that prior to the time-resolved experiments we always performed the initialization procedure that prepared the magnetization in a state that was the closest to the [010] crystallographic direction. In the case of sample $A$, the position of the easy axis – i.e, the position of the minimum of $F$ without the external field – is given mainly by the cubic anisotropy [see Fig. 1(c)]. Consequently, the application of $H_{ext}$ deepens even more this energy minimum and, therefore, reduces the precession amplitude [see Fig. 4(a)]. The precession amplitude for the highest values of $H_{ext}$ is nearly zero in this sample because at these field levels the position of the energy minimum is given solely by the direction of $H_{ext}$, which does not depend at all on any change induced in the sample by the laser pulses. In sample $B$ the uniaxial anisotropy dominates [see Fig. 1(d)] and the magnetic field again deepens the energy minimum due to the cubic anisotropy along the [010] direction. The measured non-monotonous dependence of the precession amplitude on $H_{ext}$ [see Fig. 4(a)] is a signature that for weak magnetic fields (up to ≈ 80 mT) the sensitivity of the energy minimum position to the laser-induced changes of the cubic and uniaxial anisotropies is increasing. For larger $H_{ext}$ the precession amplitude is decreasing for the same reason as in the sample $A$.

The application of $H_{ext}$ leads to the monotonous increase of the precession frequency $\omega$ – see Fig. 4(b) – that is in accord with the predictions of the classical gyromagnetic theory.[14] The measured data can be fitted by the equation[12]

$$\omega = \gamma \sqrt{(H_{ext} + H_{4\parallel} + 4\pi M_{eff} + H_{2\parallel}/2)(H_{ext} + H_{4\parallel})} \qquad (2)$$

where $4\pi M_{eff} = 4\pi M - H_{2\perp}$, and $H_{4\parallel}$, $H_{2\parallel}$ and $H_{2\perp}$ are the in-plane cubic, in-plane uniaxial and perpendicular uniaxial anisotropy fields, respectively, and $4\pi M$ represents the demagnetization term.[14] Due to the high number of parameters, the fitting of the measured data by equation (2) determines the corresponding anisotropy constants ambiguously. In order to get the correct value of the anisotropy constants, we had to take into account also the magnetization easy axis position that was determined by SQUID (see Fig. 1(c) and 1(d)). In the fitting, we used the gyromagnetic ratio $\gamma = 2.2 \times 10^5$ mA$^{-1}$s$^{-1}$, which corresponds to Mn g-factor of 2, and we obtained $4\pi M_{eff} = 236$ mT, $H_{2\parallel} = 65$ mT, $H_{4\parallel} = 98$ mT for sample $A$ and $4\pi M_{eff} = 266$ mT, $H_{2\parallel} = 60$ mT, $H_{4\parallel} = 64$ mT for sample $B$. Equation (2) also explains why the precession frequency in samples $A$ and $B$ is quite distinct at 0 mT (where the corresponding frequency difference is ≈ 30%) but nearly the same at 550 mT (where the frequency difference is ≈ 1%). Without external magnetic field, $\omega$ is given solely by the anisotropy constants, which are quite distinct in these two samples, but for higher values of $H_{ext}$ the influence of the samples magnetic anisotropy on $\omega$ is strongly suppressed.

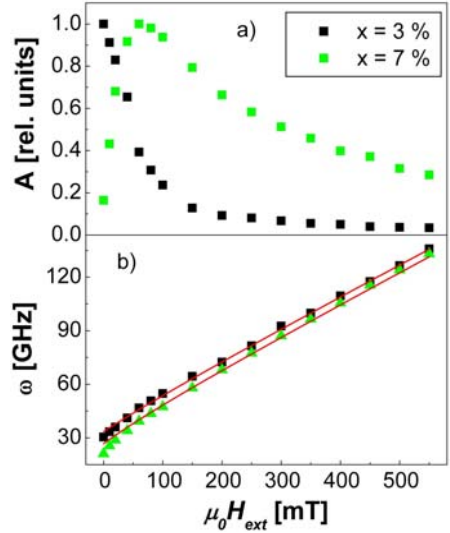

Fig. 4: Magnetic field dependence of (a) normalized precession amplitude *A* and (b) precession frequency $\omega$ (points). The lines in (b) are fits by Eq. (2) with parameters described in the main text.

In conclusion, we studied the laser-induced magnetization precession in (Ga,Mn)As. We concentrated on two samples with a different *Mn* doping that have quite distinct magnetic anisotropy. We showed that the external magnetic field strongly influences the precession of magnetization. If the magnetic field is applied close to the magnetically easy direction, the precession amplitude is quenched by the field. On the other hand, if the field is applied in the magnetically hard direction, the precession amplitude can be strongly enhanced. The anisotropy fields of the material can be deduced from the magnetic field dependence of the precession frequency.

This work was supported by Grant Agency of the Czech Republic (grant nos. P204/12/0853 and 202/09/H041), Grant Agency of the Charles University in Prague (grant nos. 443011 and SVV-2012-265306), EU ERC Advanced Grant No. 268066 and FP7-215368 SemiSpinNet, and by Preamium Academiae from the Academy of Sciences of the Czech Republic.

References
1. T. Dietl, H. Ohno, and F. Matsukura, *Phys. Rev. B* 63, 195205 (**2001**).
2. T. Jungwirth, J. Sinova, J. Mašek, J. Kučera, and A. H. MacDonald, *Rev. Mod. Phys* 78, 809 (**2006**).
3. N. Samarth, *Nature Materials* 9, 955 (**2010**).
4. J. Wang, I. Cotoros, K. M. Dani, X. Liu, J. K. Furdyna, and D. S. Chemla, *Phys. Rev. Lett.* 98, 217401 (**2007**).
5. A. Oiwa, H. Takechi, and H. Munekata, *J. Supercond. Novel Mag.* 18, 9-13 (**2005**).
6. J. Qi, Y. Xu, N. H. Tolk, X. Liu, J. K. Furdyna, and I. E. Perakis, *Appl. Phys. Lett.* 91, 112506 (**2007**).
7. Y. Hashimoto, S. Kobayashi, and H. Munekata, *Phys. Rev. Lett* 100, 067202 (**2008**).
8. E. Rozkotová, P. Němec, P. Horodyská, D. Sprinzl, F. Trojánek, P. Malý, V. Novák, K. Olejník, M. Cukr, and T. Jungwirth, *Appl. Phys. Lett.* 92, 122507 (**2008**), arXiv: 0802.2043.
9. J. Qi, Y. Xu, A. Steigerwald, X. Liu, J. K. Furdyna, I. E. Perakis, and N. H. Tolk, *Phys. Rev. B* 79, 085304 (**2009**).
10. S. Kobayashi, Y. Hashimoto, and H. Munekata, *J. of Appl. Phys.* 105, 07C519 (**2009**).

11. P. Němec, E. Rozkotová, N. Tesařová, F. Trojánek, E. De Ranieri, K. Olejník, J. Zemen, V. Novák, M. Cukr, P. Malý, and T. Jungwirth, submitted, arXiv: 1101.1049.
12. E. Rozkotová, P. Němec, N. Tesařová, P. Malý, V. Novák, K. Olejník, M. Cukr, and T. Jungwirth, *Appl. Phys. Lett.* 93, 232505 (**2008**), arXiv: 0808.3738.
13. T. Jungwirth, P. Horodyská, N. Tesařová, P. Němec, J. Šubrt, P. Malý, P. Kužel, C. Kadlec, J. Mašek, I. Němec, M. Orlita, V. Novák, K. Olejník, Z. Šobáň, P. Vašek, P. Svoboda, and J. Sinova, *Phys. Rev. Lett.* 105, 227201 (**2010**) and the Supplementary material, arXiv: 1007.4708.
14. X. Liu, and J. Furdyna, *J. Phys.: Condens. Matter* 18, R245-R279 (**2006**).
15. J. Zemen, J. Kučera, K. Olejník, and T. Jungwirth, *Phys. Rev. B* 80, 155203 (**2009**).
16. K. Hamaya, T. Watanbe, T. Taniyama, A. Oiwa, Y. Kitamoto, and Y.Yamazaki, *Phys. Rev. B* 74, 045201 (**2006**).